# Globally and Locally Optimized Pannini Projection for High FoV Rendering of 360° Images

**Falah Jabar, João Ascenso, and Maria Paula Queluz**

Instituto Superior Técnico, Universidade de Lisboa - Instituto de Telecomunicações, Av. Rovisco Pais, 1040-001 Lisboa, Portugal

Corresponding author: Maria Paula Queluz (e-mail: paula.queluz@lx.it.pt).

**Abstract**

To render a spherical (360º or omnidirectional) image on planar displays, a 2D image - called as viewport - must be obtained by projecting a sphere region on a plane, according to the user's viewing direction and a predefined field of view (FoV). However, any sphere to plan projection introduces geometric distortions, such as object stretching and/or bending of straight lines, which intensity increases with the considered FoV. In this paper, a fully automatic content-aware projection is proposed, aiming to reduce the geometric distortions when high FoVs are used. This new projection is based on the Pannini projection, whose parameters are firstly globally optimized according to the image content, followed by a local conformality improvement of relevant viewport objects. A crowdsourcing subjective test showed that the proposed projection is the most preferred solution among the considered state-of-the-art sphere to plan projections, producing viewports with a more pleasant visual quality.

## 1 Introduction

In the last few years, the interest in omnidirectional visual content has been increasing rapidly. This type of content is typically acquired with a 360º camera that often covers the whole 360º (horizontal) × 180º (vertical) viewing range. An immersive visual experience can be offered since users can interact with the content according to, at least, three degrees of freedom, corresponding to the three rotation angles (yaw, pitch, roll) around the 3D Cartesian axes. Nowadays, omnidirectional content powers a rich set of virtual reality (VR) and augmented reality (AR) applications and services in the fields of entertainment, education, medicine, arts, tourism, and sports, among others. Omnidirectional content is usually consumed by users using different types of displays, such as head-mounted displays (HMDs), mobile devices (smartphones or tablets), and standard computer monitors. Although the users can have a better immersive experience using HMD, sometimes it is not convenient or affordable and thus, smartphones or computer monitors are also rather popular. In this context, the user's quality of experience (QoE) should be maximized for all of these applications that are often deployed in a multitude of displays.

Regardless of the display device, the users only see a portion of the entire sphere - aka *viewport* - at a time. The viewport is a 2D image obtained by projecting the sphere's regions observed by the users on a plane; its content is defined by the viewing direction (VD), the horizontal field of view (HFoV), and the vertical field of view (VFoV). Naturally, the viewport contains more visual content when a large FoV is used. As shown by several studies (e.g., [5,12,17,26,27]), using a large FoV provides a more pleasant (and immersive) visual experience to users. Recently, several VR applications based on omnidirectional images are targeting a FoV close to the human FoV (e.g., [30,31]), aiming to improve the users QoE; for the horizontal direction, the human FoV is on the range 200°-220° for monocular vision, and around 114° for binocular vision; the vertical FoV is on the range 130°-135°.

Since a sphere is not a developable surface, the sphere to plan projection (SPP) needed to produce the viewport image introduces geometric distortions, such as stretching of objects and/or bending of straight lines, which intensity increases with the considered FoV. The most commonly used SPP for viewport rendering are the rectilinear, the stereographic, and the Pannini projections [1,7,13]. The rectilinear projection keeps the straightness of the lines that are also straight in the visual scene, but the objects close to the viewport borders are stretched, as shown in Figure 1a). In the stereographic projection, while object shapes are locally preserved, straight lines may be severely bent, and the fisheye effect becomes noticeable if a high FoV is used, as shown in Figure 1b).

The Pannini projection (PP) [18] involves a cylindrical projection, resulting that vertical lines keep their straightness, but horizontal lines are bent. To reduce the bending of horizontal lines, vertical compression can be applied, at the expense of increasing the objects stretching. Thus, a viewport with a good balance between stretching and bending can be obtained by tuning the projection parameters. Since the PP projection can preserve the object shapes better than the rectilinear, and vertical lines are straighter than in stereographic, it is more suitable for viewport rendering with a large FoV. As an example, in Figure 1c) the viewport resulting from the PP projection has less geometric distortions than those obtained with rectilinear and stereographic projections.

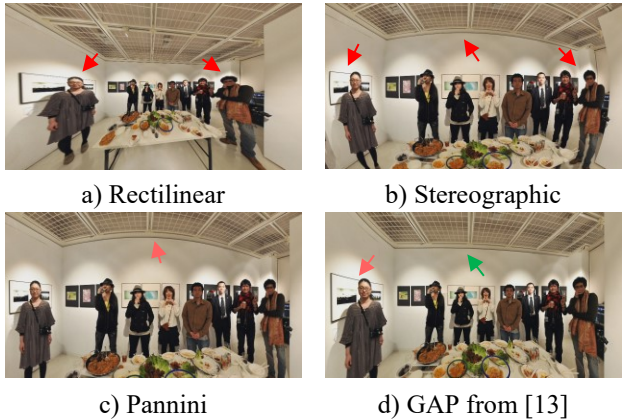

a) Rectilinear   b) Stereographic
c) Pannini   d) GAP from [13]

Figure 1: Viewport examples with a HFoV of 150° obtained with: a) Rectilinear; b) Stereographic; c) Pannini without vertical compression; c) Globally adapted Pannini (GAP) projection proposed in [13].

Recently, a few content-aware projections were developed for viewport rendering of 360º images (e.g., [13,16,32]), where the projection is globally adapted to the viewport content, i.e., the projection parameters are settled to minimize the geometric distortions for all parts of the viewport. However, these projections lack local adaptation, and thus stretching and/or bending distortions may be still visible in some image regions and structures. As an example, Figure 1d) depicts a viewport obtained with the globally adapted Pannini (GAP) projection proposed in [13]. Although it shows, globally, a lower distortion than rectilinear, stereographic and PP with fixed parameters, the stretching of the woman on the left side is clearly visible. In other contributions [3,19], some local adaptation is provided but requires the manual adjustment of parameters, or the manual identification of some image structures (e.g., lines or objects) that require some correction.

In this paper, a fully automatic content-aware projection is proposed, aiming to reduce the geometric distortions on the viewport rendering of 360º images, when high FoVs (i.e., larger than 110º) are used. As key novelty, it minimizes the viewport geometric distortions by globally and locally adapting the projection to the image content. The main contributions of this paper can be summarized as follows:

- Object-level geometric distortion correction: Using an omnidirectional semantic segmentation framework based on deep learning, objects on the visual scene are identified, and geometric distortions are minimized for the identified objects.

- Global and local optimization: A two-step procedure to minimize geometric distortions, first globally, based on the optimization of the Pannini projection, and then locally for some regions (e.g, objects) using a content-aware mesh optimization, is proposed.

- Large FoV: The proposed projection is fully automatic and content-aware, allowing to generate large FoV viewports with enhanced quality and thus increasing the user's QoE.

- Subjective assessment procedure: Often, these projections are not subjectively assessed with a well-known and already validated methodology. In this case, the evaluation of the proposed solution was performed with a crowdsourcing based pairwise comparison methodology using best practices from relevant literature, such as P.910 [11].

It is worthy to note that several applications, notably in photography, may benefit from projections covering large FoVs, allowing the rendering of panoramas, wide angle, or mosaic images. Yet, there is no automatic projection solution allowing the rendering of such wide FoV images, without visible geometric distortions. While those applications are not the focus of this paper, the proposed method may still be applied on those cases.

The rest of the paper is organized as follows: Section 2 reviews related works. Section 3 describes the proposed globally and locally adapted Pannini (GLAP) projection. Section 4 presents the GLAP parameters selection, and its performance evaluation is conducted in Section 5. Section 6 concludes the paper.

## 2 Related Work

Several sphere to plane projections were developed in the past to map wide-angle, panoramic, or 360º images to a plane. In [15], the generalized perspective projection (GPP) with varying projection center was used for the viewport rendering of 360º images. The projection includes rectilinear, stereographic, and the projections in-between. Moreover, it was shown that a trade-off between stretching and bending distortions can be obtained by changing the projection center. In [16], the content-aware generalized perspective projection (CA-GPP) was proposed, aiming to reduce, globally, the geometric distortions in the viewport. The projection center is automatically optimized based on a set of geometric distortion measures that are computed for the entire viewport. Experimental results, based on users' opinion, showed that CA-GPP leads to more pleasant viewports (and thus have higher quality) compared the rectilinear and stereographic ones. However, the CA-GPP lacks local adaptions to the viewport content and thus stretching and/or bending distortions are visible for some viewport regions, notably when high FoVs are used.

In [18], the Pannini projection (PP) was proposed to map wide-angle images. As depicted in Figure 2, this projection is accomplished in two steps: the sphere surface is first projected on an intermediate cylindrical surface, using a rectilinear projection (red lines in Figure 2); the cylinder surface is then projected onto the plane using a perspective projection with center $d$ (blue lines in Figure 2). To reduce the bending of horizontal lines, vertical compression can be applied. The forward projection equations for the Pannini projection are given by [18]:

$$x_p = S \sin(\phi), \quad (1)$$

$$y_p = (1 - vc)\,(S \tan(\theta)) + vc \left(\frac{\tan(\theta)}{\cos(\phi)}\right), \quad (2)$$

with $\quad S = \frac{d+1}{d+\cos(\phi)}, \quad (3)$

where $(\phi, \theta)$ denote, respectively, the longitude and latitude coordinates of a point on the viewing sphere, $vc$ is the vertical compression strength, and $(x_p, y_p)$ are the Cartesian coordinates of the projected point (in length units); these coordinates have their origin at the center of the viewport plane, at the tangency between this plane and the sphere. The value of $d$ allows the projection to vary its main characteristics, from rectilinear ($d = 0$) to quasi-stereographic ($d = 1$). In fact, for $d=0$ the blue and red lines of Figure 2 coincide with each other and correspond to the rectilinear projection line; for $d=1$ the blue line is close to the stereographic projection line.

In [32], an automatic optimized Pannini (OP) projection was proposed for the viewport rendering of 360º images, aiming to minimize, globally, the viewport geometric distortions. The best parameters $d$ and $vc$ are obtained automatically based on a set of global geometric distortion metrics computed on the viewport. However, the crowdsourcing based subjective evaluation showed that, in several cases, OP does not outperform the standard PP with fixed parameters; like CA-GPP, this projection lacks a local adaptation to the viewport content. Also in [32], and to improve the performance of OP, a multiple optimized Pannini (MOP) was proposed. In MOP, several optimized Pannini projections are aligned and fused to further reduce the geometric distortions. However, some viewport regions, like those containing long linear structures, may be strongly distorted, notably when different projections are used over them.

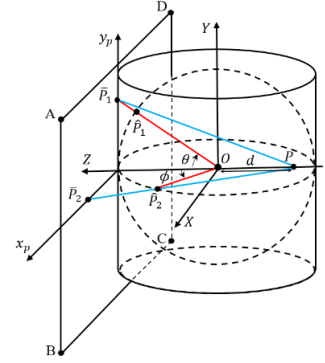

Figure 2: Pannini projection of two points, $\widehat{P}_1$ and $\widehat{P}_2$. The red lines project the points from the sphere surface to the cylinder surface; the blue lines project the points from the cylinder surface to the plane.

In [13], a globally adapted Pannini (GAP) projection was proposed to minimize, globally, the viewport geometric distortions. The parameters $d$ and $vc$ were optimized automatically based on a new set of global geometric distortion measures. The projection qualitative evaluation showed that the GAP results in viewports with better visual quality than the considered benchmark projections, including OP and MOP. However, similarly to CA-GPP and OP, GAP also lacks a local adaptation, and thus geometric distortions may be still visible for some viewport regions.

In [29], a method to correct perspective projection distortions (mostly stretching) on human faces was proposed for wide-angle photos taken from a mobile device and with FoV up to 120º. The stereographic projection is used on facial regions, which is seamlessly integrated with the rectilinear projection used for the background. However, correction of the human face geometric distortions without the rest of the body creates additional visual artifacts. Furthermore, the distortions of other objects in the scene were not considered. In addition, this projection is not suitable for omnidirectional images or when FoVs larger than 120º are used. Some other content-aware projections were developed for wide-angle or panoramic images, such as [3,19,39]; however, they are not fully automatic, requiring user interaction.

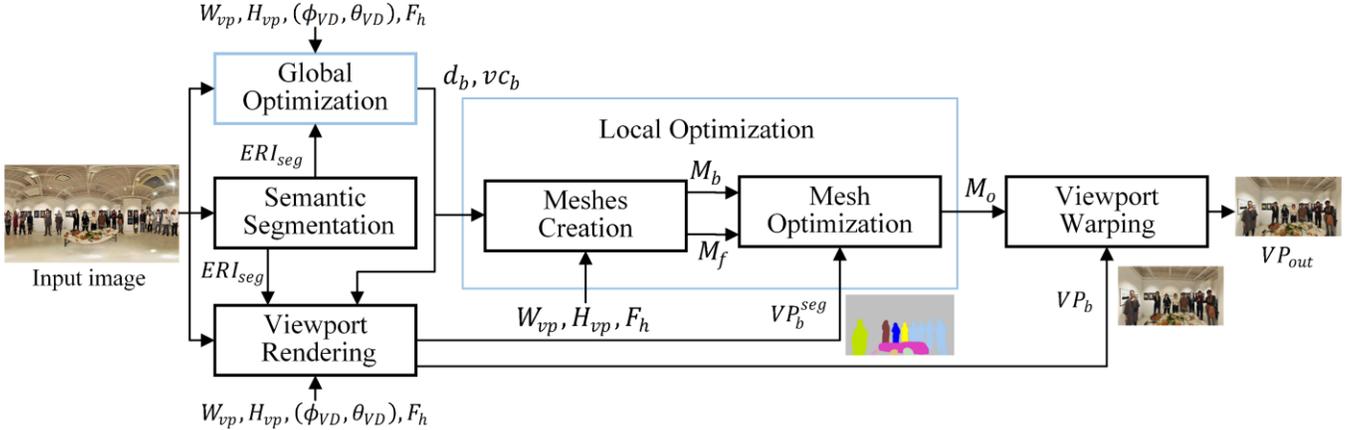

Figure 3: Proposed globally and locally adapted Pannini (GLAP) projection framework.

This paper proposes a novel projection that performs global and local adaptations (GLAP) to the content. This allows to reduce annoying geometric distortions, notably on regions where the human perception is more sensitive, such as objects. This procedure does not require any user intervention and targets the relevant case of omnidirectional images rendering.

## 3 The GLAP Projection

Figure 3 depicts the GLAP projection framework. To minimize the viewport distortions when high FoVs are used, this projection is globally and locally adapted to the viewport content, according to two optimization steps:

**1) Global optimization** – The Pannini projection is optimized considering the whole viewport, resulting in the projection parameters $(d_b, vc_b)$ with the best compromise between stretching and bending. Due to the higher visual impact of lines bending, the optimization procedure gives more importance to this distortion.

**2) Local optimization** – The projection resulting from the previous step is further improved for relevant objects. This is obtained by defining two meshes, $M_b$ and $M_f$, on the viewport plane, that are iteratively combined in one optimized mesh, $M_o$, as suggested in [29]. The indices $b$ and $f$ in $M$ stand for background and foreground, respectively. Generically, a mesh ($M$) is a grid-like structure superimposed on the viewport image, where the intersections of the grid lines constitute the vertex set $\{v_i\}$, where $i$ are the indices of the 2D coordinates on the grid of $M$, and $v_i$ denotes a 2D coordinate. The vertices are used to define how the image will be transformed. While $M_b$ corresponds to the globally optimized projection, that should be mainly applied over the background, $M_f$ corresponds to a conformal (or quasi conformal) projection, that should be mainly applied over the foreground. The goal is to increase the conformality of the foreground objects, using $M_f$, while assuring a seamless transition to $M_b$, which is mainly applied over the background.

Both optimization procedures require the detection of relevant objects, which is accomplished by the semantic segmentation block, producing a segmentation map, $ERI_{seg}$, of the input image. The Pannini projection with the globally optimized parameters, $(d_b, vc_b)$, is applied to the input image and to $ERI_{seg}$ producing, respectively, a viewport image denoted as $VP_b$, and its corresponding segmentation map, denoted as $VP_b^{seg}$, which is used by the mesh optimization procedure. Finally, $VP_b$ is warped according to the optimized mesh, $M_o$, to obtain the final output viewport, $VP_{out}$. The main steps involved in the GLAP projection are detailed in the following sections.

### 3.1 Semantic Segmentation

Semantic segmentation refers to the task of assigning a class label (e.g., people, chair, car, etc.) to objects in the image; objects in the same class have the same label. It is widely used in computer vision tasks, using typically 2D images. In [35–38], several semantic segmentation models were also proposed for panoramic or 360º images, targeting autonomous driving, but only outdoor images were considered. In this paper, semantic segmentation is obtained for both indoor and outdoor images by transforming the input image, in equirectangular format (ERI), to cubic format [33]. This results in six 2D images (the cube faces), each one having a horizontal and vertical FoV of 90º. A high performance semantic segmentation model – the Auto-DeepLab – proposed in [21], is then applied to each cube face. This model uses multi-scale inference and the network backbone Xception-65, it was pre-trained on ImageNet [28] and on MS-COCO [20] datasets, and trained on the PASCAL VOC 2012 dataset [4], which contains 20 foreground object classes and one background class. The segmentation of all six cube face images is transformed back to equirectangular format. Since objects in the same class have the same label, disconnected objects inside the same class are differentiated from each other by applying the connected component analysis (CCA) [22], with 4-connectivity. Figure 4 depicts an example of an *ERI* image, its semantic segmentation using Auto-DeepLab, and the resulting *ERI* segmentation map ($ERI_{seg}$) after CCA, where disconnected objects are represented with different colors.

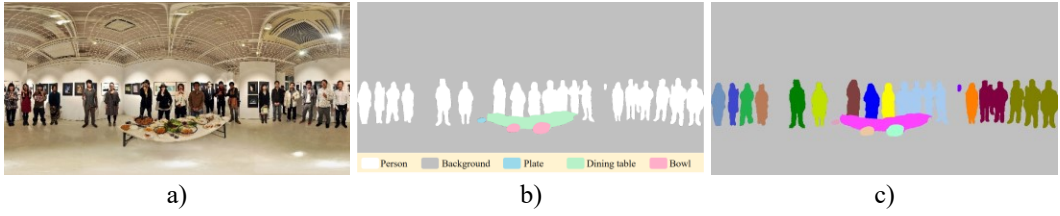

Figure 4: a) Example of an ERI image; b) Its semantic segmentation; c) Its final segmentation map, ERI_seg.

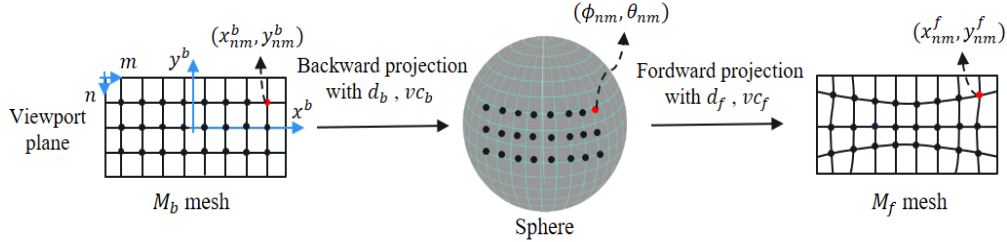

Figure 5: $M_b$ and $M_f$ meshes generation procedure.

### 3.2  Global Optimization

The global optimization aims to find out the Pannini projection parameters, $(d_b, vc_b)$, that result in the best compromise between objects stretching and lines bending, for a viewport rendered according to the user viewing direction, $(\phi_{VD}, \theta_{VD})$, and with a predefined horizontal field of view, $F_h$, and spatial resolution $(W_{vp}, H_{vp})$. The spherical coordinates $\phi$ and $\theta$ denote, respectively, the longitude and the latitude on the sphere. In this procedure, the best - in the perceived quality sense - projection parameters, $(d_b, vc_b)$, are obtained by minimizing a simple cost function:

$$(d_b, vc_b) = \underset{(d,vc)}{\mathrm{argmin}}\big(\beta\, S(d, vc) + B(d, vc)\big), \quad (4)$$

varying $d$ in the range [0.1,1] and $vc$ in the range [0,1], with steps $\Delta d = \Delta vc = 0.1$. $S(d, vc)$ and $B(d, vc)$ are, respectively, the resulting viewport stretching and bending measures for projection parameters $(d, vc)$, both measures normalized to the interval [0,1], and $\beta$ is the stretching to bending ratio. The viewport stretching was computed using the object based Tissot area distortion measure ($OAD^{TO}$) proposed in [13], while the viewport bending was obtained using the Line Measure Combination (*LMC*), proposed in [14]. The value of $\beta$ (set to 0.17) was obtained through the subjective assessment of several viewports, rendered for different values of $(d, vc)$. In [13], $\beta$ was obtained for the GAP projection seeking the best balance between bending and stretching distortions. However, in this work, to preserve the straightness of the lines as much as possible, more importance was given to the line bending than to the stretching of the objects (and since the local optimization will only improve the latter).

### 3.3  Meshes Creation

Two meshes, $M_b$ and $M_f$, are generated on the viewport plane, as depicted in Figure 5, using the Pannini backward and forward projections. This allows to obtain, for a given position in the viewport, the corresponding positions in both $M_b$ and $M_f$:

• $M_b$ **mesh creation** - A uniform grid mesh, $M_b = \{b_i\}$, is defined over $VP_b$, consisting of a vertex set $\{b_i\}$, where $b_i$ refers to the *i*-th vertex Cartesian coordinates, $(x^b, y^b)$, in length units, with origin at the center of the viewport plane. For a given integer position of $VP_b$, $(n, m)$, with a coordinate system centered on the top-left corner of the viewport plane, the corresponding Cartesian coordinates, $(x^b_{nm}, y^b_{nm})$, can be computed by:

$$x^b_{nm} = 2\,(d_b + 1)\frac{\sin\!\left(\frac{F_h}{2}\right)}{d_b + \cos\!\left(\frac{F_h}{2}\right)}\left(\frac{m + 0.5}{W_m} - \frac{1}{2}\right), 0 \le m < W_m \quad (5)$$

$$y^b_{nm} = 2\,\tan\!\left(\frac{F_v}{2}\right)\left(\frac{1}{2} - \frac{n + 0.5}{H_m}\right), 0 \le n < H_m \quad (6)$$

where $F_h$ is horizontal FoV and $F_v$ is the vertical FoV, related by:

$$F_v = 2\,\tan^{-1}\left(\frac{(d_b+1)\sin\left(\frac{F_h}{2}\right)}{AR\left(d_b+\cos\left(\frac{F_h}{2}\right)\right)}\right), \quad (7)$$

and $W_m$ and $H_m$ are the horizontal and vertical mesh resolution, respectively, which were set to $W_m = W_{vp}/c$ and $H_m = H_{vp}/c$, being $c$ a constant; $AR$ is the viewport aspect ratio.

• **$M_f$ mesh creation** - The $M_b$ mesh coordinates, $(x_{nm}^b, y_{nm}^b)$, are projected back to the sphere, using the Pannini backward projection with parameters $(d_b, vc_b)$, resulting in the corresponding spherical coordinates $(\phi_{nm}, \theta_{nm})$. These are then projected to the plane using the Pannini forward projection with parameters $(d_f, vc_f)$, resulting in the corresponding $M_f$ mesh coordinates, $(x_{nm}^f, y_{nm}^f)$, of a vertex set $\{\boldsymbol{f}_i\}$. Thus, $M_f$ represents the initial viewport reprojected according to $(d_f, vc_f)$, that should preserve the objects conformality (e.g., stereographic Pannini ($d_f = 1, vc_f = 0$)). The selection of these parameters is detailed in Section 4.

Note that to get a uniform $M_f$ mesh with the same resolution as $M_b$, the procedure was implemented in the other way around: for each integer position $(n, m)$ associated with a vertex $\boldsymbol{f}_i$, the corresponding Cartesian coordinates, $(x_{nm}^f, y_{nm}^f)$, were obtained by first back projecting to the sphere with $(d_f, vc_f)$, and then forward projecting to the viewport plan with $(d_b, vc_b)$. The pseudocode explaining this procedure is provided below.

**Algorithm: Pannini meshes creation**
1: Input: $d_f, vc_f, d_b, vc_b, W_m, H_m, F_h$
2: Output: $M_b, M_f$
3:   for $n = 1$ to $H_m$
4:     for $m = 1$ to $W_m$
5:       compute $x_{nm}^b, y_{nm}^b$ using (2) to (4)
6:       compute $(\phi_{nm}, \theta_{nm})$ using (4) to (7) in [13] with $d_f, vc_f$
7:       compute $x_{nm}^f, y_{nm}^f$ using (1) to (3) in [13] with $d_b, vc_b$
8:     end
9:   end

### 3.4 Mesh Optimization

Based on [29], a mesh optimization algorithm is proposed which iterates between $M_b$ and $M_f$, adding smooth changes, to obtain an optimal mesh $M_o = \{\boldsymbol{o}_i\}$. This mesh has the following properties: *i)* object shapes are preserved; *ii)* straightness of background lines are preserved; *iii)* abrupt transitions at the object borders (due to the use of two different meshes) are avoided.

A mesh denoted as $M_v = \{\boldsymbol{v}_i\}$ is defined, consisting of a vertex set $\{\boldsymbol{v}_i\}$, where initially $\{\boldsymbol{v}_i\} = \{\boldsymbol{b}_i\}$. The optimized mesh results from minimizing the following cost function:

$$\{\boldsymbol{o}_i\} = \underset{\{\boldsymbol{v}_i\}}{\arg\min}\; E_t(\{\boldsymbol{v}_i\}), \quad (8)$$

where $E_t$ is a weighted sum of energy terms, expressed by:

$$E_t = \lambda_c E_c + \lambda_b E_b + \lambda_s E_s + \lambda_a E_a. \quad (9)$$

and $E_c, E_b, E_s$, and $E_a$ are, respectively, object conformality, line distortion, smoothness, and asymmetric energy terms; $\lambda_c, \lambda_b, \lambda_s$, and $\lambda_a$ are the weights for the corresponding energy terms. Each one of these energy terms is explained below:

• **Object conformality term** - For each object identified in $VP_b^{seg}$, a conformality term is computed by:

$$O_c(k) = \sum_{i \in I_k} m_i \|\boldsymbol{v}_i - \boldsymbol{f}_i\|_2^2, \quad (10)$$

where $k$ is the object index; $I_k$ is the set of vertex indices on the $k$-th object; $m_i$ is the correction strength for the $i$-th vertex; $\boldsymbol{f}_i$ is the vertex in the $M_f$ mesh; $\boldsymbol{v}_i$ is the vertex in the $M_v$ mesh; $\|.\|_2^2$ denotes the squared Euclidean distance. The $O_c$ term encourages the object regions to follow the $M_f$ mesh, where the object shapes are preserved.

Objects located at the viewport borders have higher distortions than objects close to the viewport center, and thus require more correction. To account it, the following sigmoid function is defined:

$$m_i = \frac{1}{1 + \exp\left(-\frac{r_i - r_1}{r_2}\right)}, \quad (11)$$

where $r_i$ is the radial distance of $\boldsymbol{b}_i$ from the viewport center, $r_1$ and $r_2$ are parameters controlling the attenuations of the correction strength and chosen such that $m_i = 0.01$ at the viewport center, and $m_i = 1$ at the viewport border.

The total object conformality is then computed by the sum of all objects conformality and expressed by:

$$E_c = \sum_k O_c(k),\quad (12)$$

- **Line distortion term** – To preserve the straightness of the lines on the boundaries between objects and background, where different projections are applied, the following line distortion energy term is computed:

$$E_{ld} = \sum_i \sum_{j \in N(i)} \|\boldsymbol{v}_i - \boldsymbol{v}_j \times e_{ij}\|_2^2,\quad (13)$$

where $N(.)$ represents the 4-way vertex neighborhood, $e_{ij}$ is the unit vector along the direction $\boldsymbol{b}_i - \boldsymbol{b}_j$ in the $M_b$ mesh, that preserves the lines straightness, and $\times$ denotes the cross product.

- **Smoothness term** - To have a smooth transition at the object borders, the following smoothness term is computed:

$$E_s = \sum_i \sum_{j \in N(i)} \|\boldsymbol{v}_i - \boldsymbol{v}_j\|_2^2.\quad (14)$$

This term encourages smoothness between 4-way adjacent vertices and thus avoids abrupt changes in the final viewport.

- **Asymmetric cost term** - Due to the mesh optimization that tries to satisfy the previous terms, some visual artifacts (e.g., geometric distortions and black regions) may appear at regions close to the viewport borders. Thus, to reduce these artifacts, the following asymmetric cost term is computed:

$$E_a = E_{le} + E_{ri} + E_{to} + E_{bo},\quad (15)$$

where $E_{le}, E_{ri}, E_{to},$ and $E_{bo}$ are, respectively, left, right, top, and bottom mesh boundary constraints, given by:

$$E_{le} = \frac{1}{H_m} \sum_{i \in \partial_{left}} \mathbb{I}(v_{i,x} > 0) \times \|v_{i,x}\|_2^2 \quad (16)$$

$$E_{ri} = \frac{1}{H_m} \sum_{i \in \partial_{right}} \mathbb{I}(v_{i,x} < W_m) \times \|v_{i,x} - W_m\|_2^2 \quad (17)$$

$$E_{to} = \frac{1}{W_m} \sum_{i \in \partial_{top}} \mathbb{I}(v_{i,y} > 0) \times \|v_{i,y}\|_2^2 \quad (18)$$

$$E_{bo} = \frac{1}{W_m} \sum_{i \in \partial_{bottom}} \mathbb{I}(v_{i,y} < H_m) \times \|v_{i,y} - H_m\|_2^2,\quad (19)$$

where $\mathbb{I}(.)$ is the indicator function that returns 1 for the true condition and 0 otherwise; $\partial_*$ are the original mesh boundaries; $W_m$ and $H_m$ are the horizontal and vertical mesh resolutions; $v_{i,x}, v_{i,y}$ are the $x$ and $y$ coordinates of vertex $v_i$.

A gradient-based algorithm [24], with 100 iterations and a learning rate of 0.02, was used for the mesh optimization.

This method was implemented in PyTorch [25], which is computationally efficient and suitable for mesh optimization.

### 3.5 Viewport Warping

The final viewport, $VP_{out}$, is obtained by warping the globally optimized viewport, $VP_b$, according to the optimized mesh, $M_o$. The warping package available in [23] was used for this purpose. This process requires interpolation for non-integer pixel positions. in this work, bilinear interpolation was used.

Figure 6 depicts: a viewport, $VP_b$, obtained from the globally optimized Pannini projection with $d_b = 0.5$ and $vc_b = 0.6$; its segmentation map, $VP_b^{seg}$; a viewport, $VP_f$, obtained by warping $VP_b$ according to a $M_f$ mesh generated with $d_f = 0.5$ and $vc_f = 0$; and the final optimized viewport, $VP_{out}$. As can be seen, the objects stretching presented in $VP_b$ (e.g., the girl on the left side is vertically stretched), is significantly reduced in the $VP_{out}$, while straight lines in the background remain straight. Figure 6e) shows the optical flow mask [34] overlaid on $VP_b$. This mask was computed between the two meshes, $M_b$ and $M_o$, and it highlights the $VP_b$ regions that were modified by the mesh optimization procedure. As shown in Figure 6e), the bottom-left and the bottom-right regions have the strongest flow (or projection modifications, to reduce the stretching) compared to other regions, which was expected since there are two objects (lady on the left and boy on the right) located in these regions, that were too much stretched in the vertical direction, on Figure 6a).

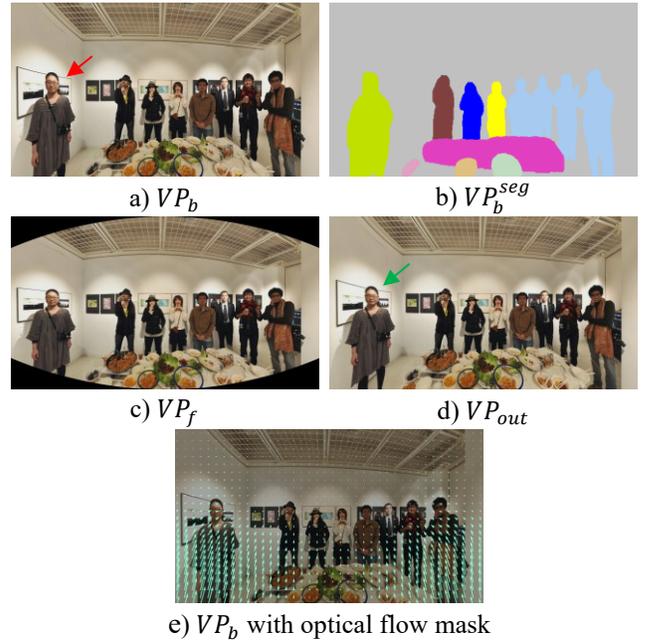

a) $VP_b$  
b) $VP_b^{seg}$  
c) $VP_f$  
d) $VP_{out}$  
e) $VP_b$ with optical flow mask

Figure 6: a) Globally optimized Pannini viewport with HFoV of 150º; b) its segmentation map; c) viewport obtained by warping $VP_b$, according to the $M_f$ mesh; d) Final output viewport; e) $VP_b$ with optical flow mask.

## 4  GLAP Projection Parameters Selection

To obtain the $M_f$ mesh, the corresponding Pannini projection parameters, $(d_f, vc_f)$, need to be found. While a stereographic Pannini ( $d_f = 1, vc_f = 0$) favours the conformality of the objects, it may result in visible distortions on the objects boundaries, if the global projection parameters have very distinct values.

The appropriate values of $(d_f, vc_f)$ were obtained by visual inspection of the optimized viewports, $VP_{out}$, for several 360º images, varying $d_f$ in the range of [0.1, 1] with a step $\Delta d = 0.1$, and $vc_f = 0$. The $vc_f$ value was set to 0 since, for $d_f \neq 0$ and $vc_f > 0$, object stretching becomes visible. It was found that if $|d_b - d_f| > 0.2$, the regions close to the object boundaries may be distorted on the final viewport, particularly if it contains straight lines. Accordingly, $d_f$ was set to $d_b + 0.2$, being $d_b$ automatically obtained by the global optimization procedure.

The cost function defined by (9) has four parameters $\lambda_c$, $\lambda_b$, $\lambda_s$, and $\lambda_a$. To tune these parameters, the following steps were applied:

1) Initialize the parameters according to $\lambda_c = 4$, $\lambda_b = 2$, $\lambda_s = 0.5$ and $\lambda_a = 4$. Although other values are possible, this initialization provided a good starting point.
2) Tune the parameters sequentially, one at a time, varying their values in the range [0.1, 6] with a step size of 0.1, and retain the value that leads to the best viewport quality by visual inspection.

These steps were applied to several 360º images, and the best values found for $\lambda_c$, $\lambda_b$, $\lambda_s$, and $\lambda_a$ were, respectively, 0.3, 1.5, 0.5, 3. To evaluate the impact of these parameters on the final output, the GLAP viewport was obtained with the tuned parameter values, being the result shown in Figure 7a). After, the projection was repeated with each parameter set to 0, one at a time, and the results are shown in Figure 7b)-e). When $\lambda_c = 0$, the objects are stretched in the output viewport (*cf.* Figure 7b). When $\lambda_b = 0$, the straight lines between the objects and background are deformed, e.g., radial lines behind the girl on the left side of Figure 7c). When $\lambda_s = 0$, dramatic changes happen for some image regions, e.g., the painting behind the girl on the left side of Figure 7d). When $\lambda_a = 0$, the regions close to the viewport borders are distorted (*cf.* Figure 7e). Figure 7f)-j) show the optical flow mask for different parameter configuration, overlaid on $VP_b$.

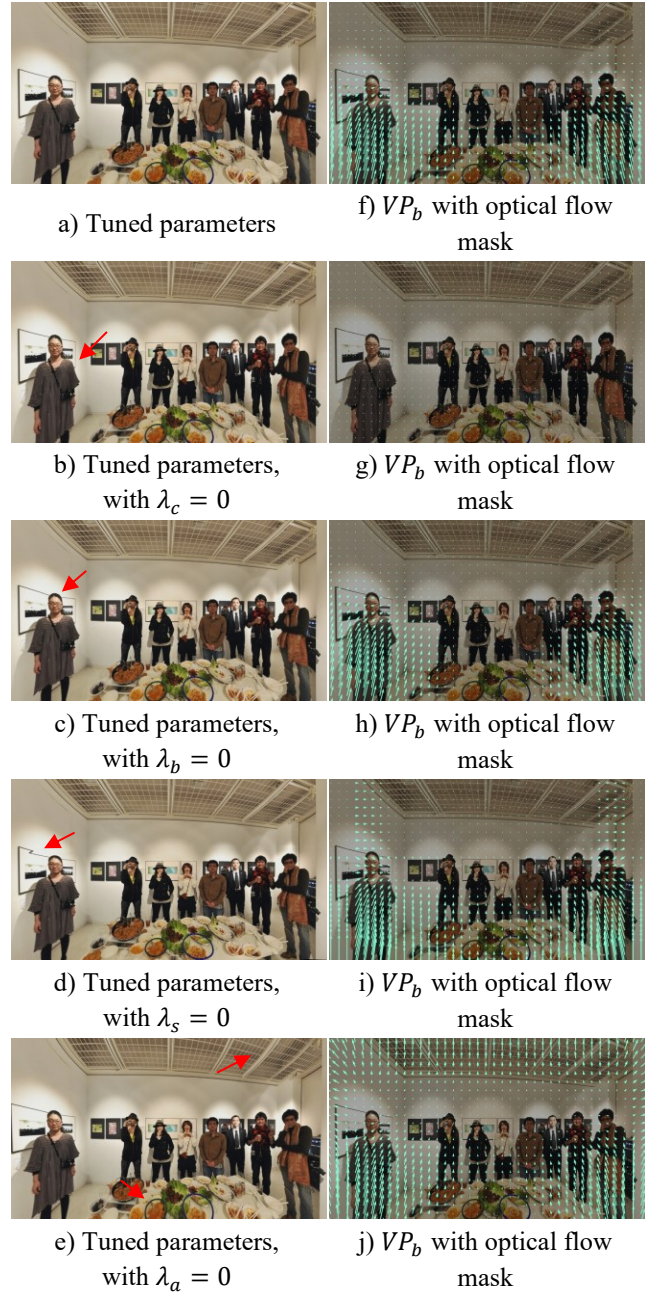

a) Tuned parameters

f) $VP_b$ with optical flow mask

b) Tuned parameters, with $\lambda_c = 0$

g) $VP_b$ with optical flow mask

c) Tuned parameters, with $\lambda_b = 0$

h) $VP_b$ with optical flow mask

d) Tuned parameters, with $\lambda_s = 0$

i) $VP_b$ with optical flow mask

e) Tuned parameters, with $\lambda_a = 0$

j) $VP_b$ with optical flow mask

Figure 7: Viewports on the left side were obtained with GLAP using a HFoV of 150º and several parameters configuration; viewports on the right side correspond to $VP_b$ with optical flow mask (in green), showing the viewport regions modified by the mesh optimization procedure.

## 5 Projection Performance Evaluation

This section describes the crowdsourcing subjective assessment evaluation of the proposed GLAP.

### 5.1 Test Conditions

The GLAP viewports were obtained with the parameter values determined in Section 4. The viewports had a HFoV of 150°, as in [32], and a spatial resolution of $1816 \times 1020$ pixels ($AR = 16/9$) as recommended in [2]. To speed up the optimization procedure, the mesh dimension was set to $181 \times 102$, which corresponds to $\lfloor \frac{W_{vp}}{10} \rfloor, \lfloor \frac{H_{vp}}{10} \rfloor$, where $\lfloor . \rfloor$ is the floor operator (recall that $W_{vp}$ and $H_{vp}$ are, respectively, the horizontal and vertical viewport spatial resolution). After optimization, the optimized mesh was resized with bilinear interpolation to the viewport resolution.

The following five benchmark projections were considered: PP with fixed parameters, ($d = 0.5, vc = 0$); GPP with fixed parameter, $d = 0.5$; OP and MOP projections proposed in [32]; and the globally adapted projection, GAP, proposed in [13]. While the PP and GPP projections are widely established content-unaware projections, the OP, GAP and MOP are the best automatic content-aware projections, all of them based on the Pannini projection. The OP and MOP viewports were obtained from the author of [32] since the source code was not available.

Eight omnidirectional images in equirectangular format (ERI) were used in the subjective assessment; they are depicted in Figure 8, and are available in [41]. To have different image content characteristics, e.g., objects near and far away from the camera and the presence or absence of people, two groups of images, *G1* and *G2* (presented in Figure 8), were taken from two different datasets: *G1* from [32] and *G2* from [6]. Per image, one viewing direction was considered. Thus, six viewports were obtained, corresponding to the proposed and the benchmark projections.

### 5.2 Subjective Evaluation Method

The pairwise comparison (PC) subjective evaluation method was used, where two images are shown side by side and the observer selects the one that has the best quality, in his opinion; this method is often used for evaluating rendering methods [8–10].

For each 360° image, a complete set of comparisons was performed (i.e., all possible pairs of comparisons), which resulted in 15 comparisons per 360º image in *G1*. However, to limit the test duration to less than half an hour, thus avoiding the observer fatigue, viewports from OP and MOP

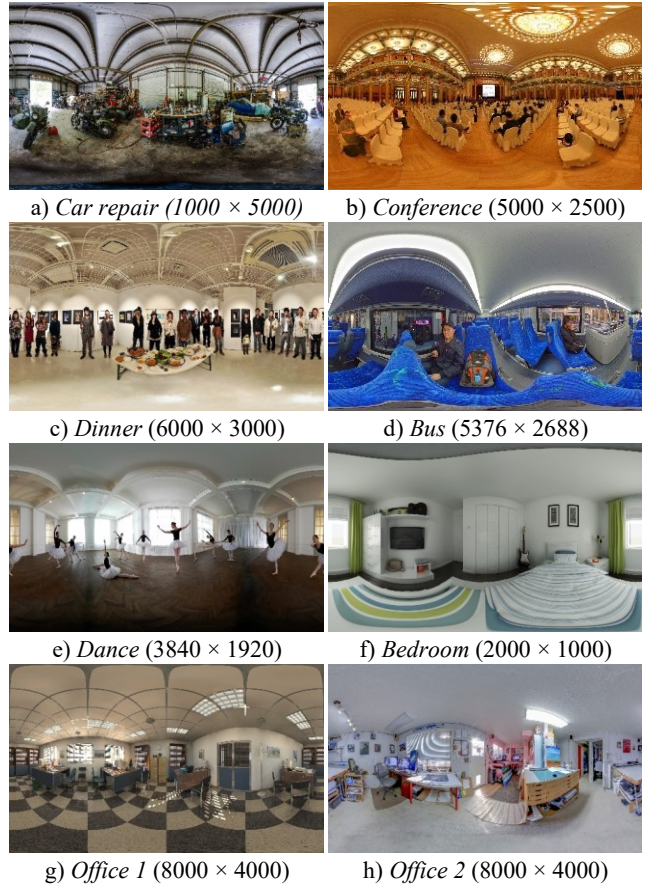

a) *Car repair (1000 × 5000)*    b) *Conference (5000 × 2500)*

c) *Dinner (6000 × 3000)*    d) *Bus (5376 × 2688)*

e) *Dance (3840 × 1920)*    f) *Bedroom (2000 × 1000)*

g) *Office 1 (8000 × 4000)*    h) *Office 2 (8000 × 4000)*

Figure 8: Omnidirectional images used in the subjective test, and their spatial resolutions.

were excluded from the test in *G2*, thus reducing to six the number of comparisons per 360º image. Table 1 presents the used 360º images, projections, and the total number of comparisons, for *G1* (15 (comparisons/image) × 4(images) = 60) and *G2* (6 (comparisons/image) × 4(images) = 24).

The subjective test was conducted online through a web-based crowdsourcing interface that allows to display two viewports, 'A' and 'B', side by side, with random order and position. The observers were asked to select the viewport, 'A' or 'B', that has the best quality in his/her opinion, or option 'A=B' in case of no difference, to avoid random preference selections. The total number of observers that participated in the online subjective test was 30. The obtained viewports and the resulting PC subjective scores are available in [41].

### 5.3 Outlier Detection

For each observer, the transitivity satisfaction rate, $R$, was computed [40]. The transitivity rule is violated when the observer's scores are inconsistent for three compared stimuli, $A, B, C$, forming a circular triad (e.g., $A > B$, $B > C$, $C > A$, where $>$ indicates "better than"). The score reliability, $R_o$, of observer o, was computed as:

$$R_o = 1 - \frac{d_o}{h_o}, \quad (20)$$

where $d_o$ is the number of detected circular triads for participant $o$ and $h_o$ is the total number of comparisons made by participant $o$. As recommended in [40], if $R_o < 0.9$ then observer $o$ is considered an outlier. Four outliers were detected, and their subjective scores were not further considered.

Table 1: The 360º images, the used projections, and the total number of comparisons for each image group.

| Group | 360º images | Dataset | Projections | Number of comparisons |
|---|---|---|---|---|
| *G1* | *Dance, Bedroom, Office 1, Office 2* | [32] | GPP, PP, OP, GAP, MOP, GLAP | 60 |
| *G2* | *Car repair, Conference, Dinner, Bus* | [6] | GPP, PP, GAP, GLAP | 24 |

Table 2: Preference probabilities for compared projections in *G1*/*G2*. NA corresponds to Not Available.

|  | GPP [15] | PP [18] | OP [32] | MOP [32] | GAP [13] | GLAP |
|---|---|---|---|---|---|---|
| GPP | - | 0.28/0.67 | 0.29/NA | 0.19/NA | 0.25/0.22 | 0.16/0.15 |
| PP | 0.72/0.33 | - | 0.39/NA | 0.56/NA | 0.30/0.23 | 0.16/0.11 |
| OP | 0.71/NA | 0.61/NA | - | 0.61/NA | 0.32/NA | 0.27/NA |
| MOP | 0.81/NA | 0.44/NA | 0.39/NA | - | 0.26/NA | 0.21/NA |
| GAP | 0.75/0.78 | 0.70/0.77 | 0.68/NA | 0.74/NA | - | 0.28/0.09 |
| GLAP | **0.84**/**0.85** | **0.84**/**0.89** | **0.73**/NA | **0.79**/NA | **0.72**/**0.91** | - |

### 5.4 Subjective Test Results and Analysis

For each compared viewport pair (A, B), the winning frequency, $w_{AB}$, which represents the number of times viewport A was preferred over viewport B, was computed. The tie cases were solved by giving a score of 0.5 to each viewport, whenever the observer selected the option 'A=B'. The probability of selecting A against $B$, is given by:

$$P_{AB} = P(A > B) = \frac{w_{AB}}{O}, \quad (21)$$

where $O$ is the total number of observers. The preference probabilities were then translated to absolute quality scores using the Bradley-Terry (BT) model, as described in [17].

Table 2 presents the preferences probabilities for the considered projections and per image group, computed by (21) and averaged over the different images in each group. In this table, the values in green and blue color correspond, respectively, to the preference probabilities for the images in *G1* and in *G2*. Accordingly, the following conclusions can be taken:

- For the images in *G1*, the proposed GLAP projection is preferred over all benchmark projections by 72% (minimum) to 84% (maximum) of the subjects. The GLAP outperforms the best content-aware benchmark projection available in the literature, MOP, by a large margin, since 79% of the subjects prefered it. For the images in *G2*, the GLAP is preferred over the GPP and PP, by 85% and 89% of the subjects, respectively.

- The GLAP is prefereed over GAP by 72% of the subjects for the images in *G1*, and by a large margin, 91%, for images in *G2*, showing the advantage of having the projection locally adapted to the content (recall that in GAP the projection is just globally optimized).

Figure 9 depicts the resulting BT scores obtained for each projection and image group. As shown, the proposed GLAP obtained the highest quality scores for all images in both *G1* and *G2* groups. Interestingly, in *G1* the benchmark projections results are not consistent and highly depend on the image content, e.g. GAP has the highest quality for images *Bedroom* and *Office 2*, while for the *Dance* image has lower quality even when compared to the content-unaware Pannini projection (PP). This behaviour does not occur for the proposed GLAP projection which has consistent quality scores for all images in both image groups. In summary, the proposed GLAP leads to higher perceived quality gains compared to previous state-of-the-art, notably content-aware projections based on the Pannini projection.

### 5.5 Projection Qualitative Evaluations

Figure 10 depicts some viewport examples obtained for the proposed GLAP and for the benchmark projections OP and MOP proposed in [32], allowing the following comparisons:

- **GLAP *vs* OP** - The GLAP viewports have less geometric distortion than the viewports resulting from OP. For *Bedroom*, the horizontal lines on the ceiling and on the floor are straighter for GLAP. In *Office 1*, the chair on the left side is more conformal and the horizontal line on the ceiling is straighter for GLAP. In *Office 2*, the monitor and the chair on the left side are stretched too much for OP. In *Furniture,* GLAP kept the horizontal lines as straight as OP, but the objects shape (e.g., the table and the chairs on the right side) is more conformal for GLAP.

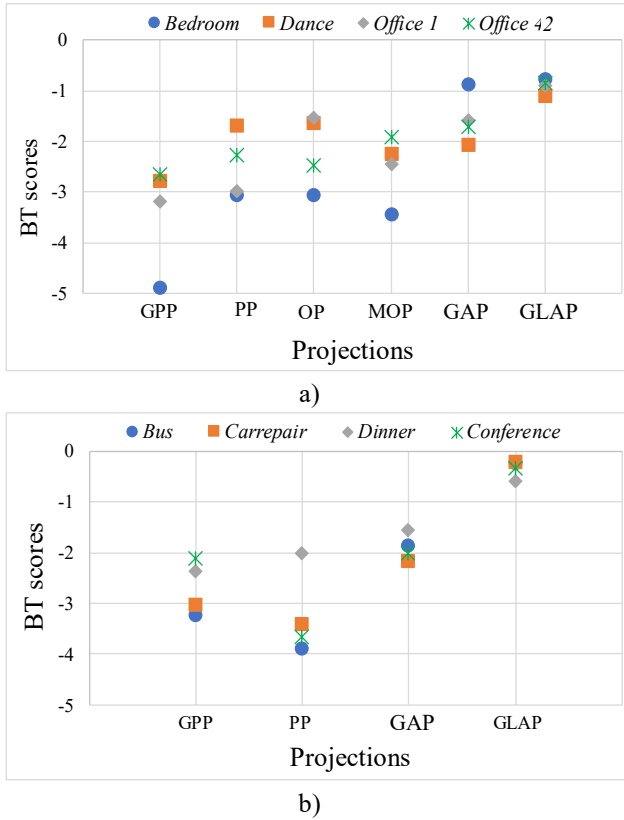

Figure 9: BT scores *vs*. projections for each considered 360º image in a) *G1* and b) *G2*.

- **GLAP *vs* MOP -** The viewports obtained for MOP have more geometric distortions than the viewports resulting from GLAP. MOP has a poor balance between bending and stretching; the horizontal lines are too much bent, and some vertical lines are also bent for some images, e.g., in *Furniture*. Also, in *Furniture*, the table on the right side is globally deformed.

Figure 11 depicts examples of viewports obtained for the benchmark projection GAP [13] and proposed GLAP. For this evaluation, the locally optimized projection (LOP) proposed in [29] was also considered, but for correcting general objects (and not just human faces, as in [29]). Figure 12 depicts the same viewports of Figure 11 but with cropped objects to better compare their conformality for different projections. The following conclusions can be taken:

- **GLAP *vs* LOP -** The GLAP viewports have a much better perceived quality than LOP viewports. The LOP stretches the objects too much since it uses a mixture of two projections, rectilinear and stereographic, and the former is known for a strong perspective effect and objects stretching, notably when a large FoV is used.
- **GLAP *vs* GAP -** The horizontal lines are less bent for the GLAP, particularly for the *Conference*, *Carrepair*, and *Bus* viewports. The stretching distortion, that is visible for some objects/regions, are significantly reduced in the GLAP viewports.

## 6 Conclusion

This paper proposes a fully automatic Pannini-based projection for the viewport rendering of 360° images. The projection is globally and locally adapted to the viewport content, and is able to produce high FoV viewports with significantly better visual quality than the best state-of-the-art benchmark projections. This allows to enhance the user's QoE for several applications and services that make use of 360° images (e.g., VR and AR applications).

An interesting direction for future work is dynamic projection optimization under user navigation. Depending on the omnidirectional image content and viewing direction, the best projection could vary. Naturally, changes in the projection should be smooth and always aiming to improve the perceived quality or immersiveness. This may require additional and specific subjective tests where user interaction is allowed or simulated. This dynamic projection optimization may also be useful for omnidirectional video rendering. In this case, other factors may have an impact on the optimization, such as objects motion, camera motion and the existence of scene cuts.

Another interesting direction for future work is projection optimization for HMD devices. It is expected that the geometric distortions impact will be not the same for HMD compared to computer monitors. The HMDs have two displays that are closer to the user's eyes, and thus the users may pay more attention to the regions near the point of fixation (foveated vision), compared to the regions away from that point (peripheral vision). Therefore, eye-tracking technology could be integrated in the proposed viewport projection. Naturally, additional subjective evaluations using HMD may be required to validate the proposed projection.

**Declaration of competing interest**

The authors declare that they have no known competing financial interests or personal relationships that could have appeared to influence the work reported in this paper.

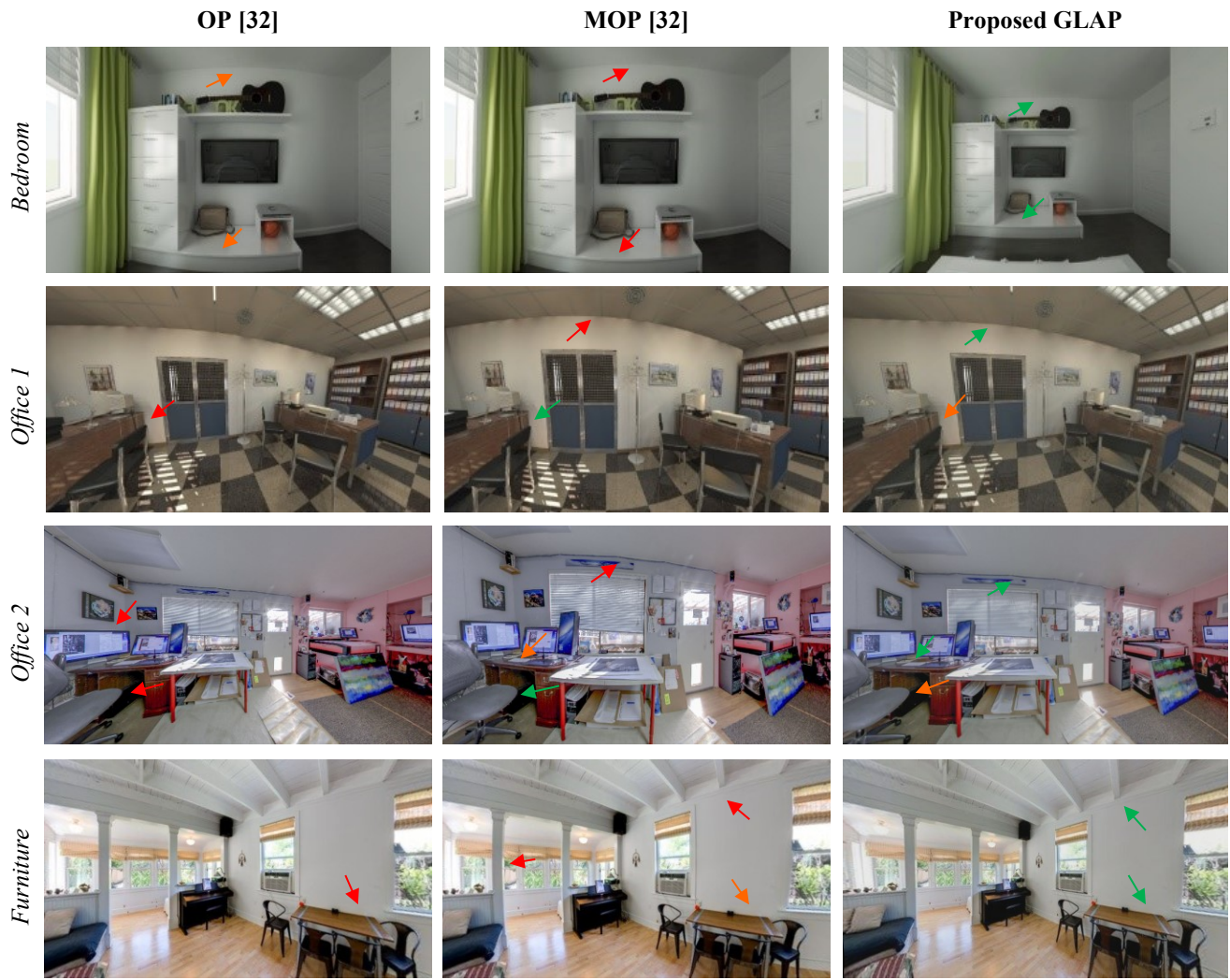

Figure 10: Viewport examples obtained with OP and MOP from [32], and with proposed GLAP, using a HFoV of 150º. The red, orange, and green arrows indicate, respectively, the objects/regions with high, medium, and low geometric distortions.

**Data availability**

The used omnidirectional images, rendered viewports, and the resulting PC subjective scores are available in: https://www.dropbox.com/sh/8eloomksm9itdn2/AAAaFMVib09Ol5nHImRv6QUga?dl=0

**Acknowledgments**

This work was funded by FCT/MCTES through national funds and when applicable co-funded by EU funds under the project UIDB/50008/2020.

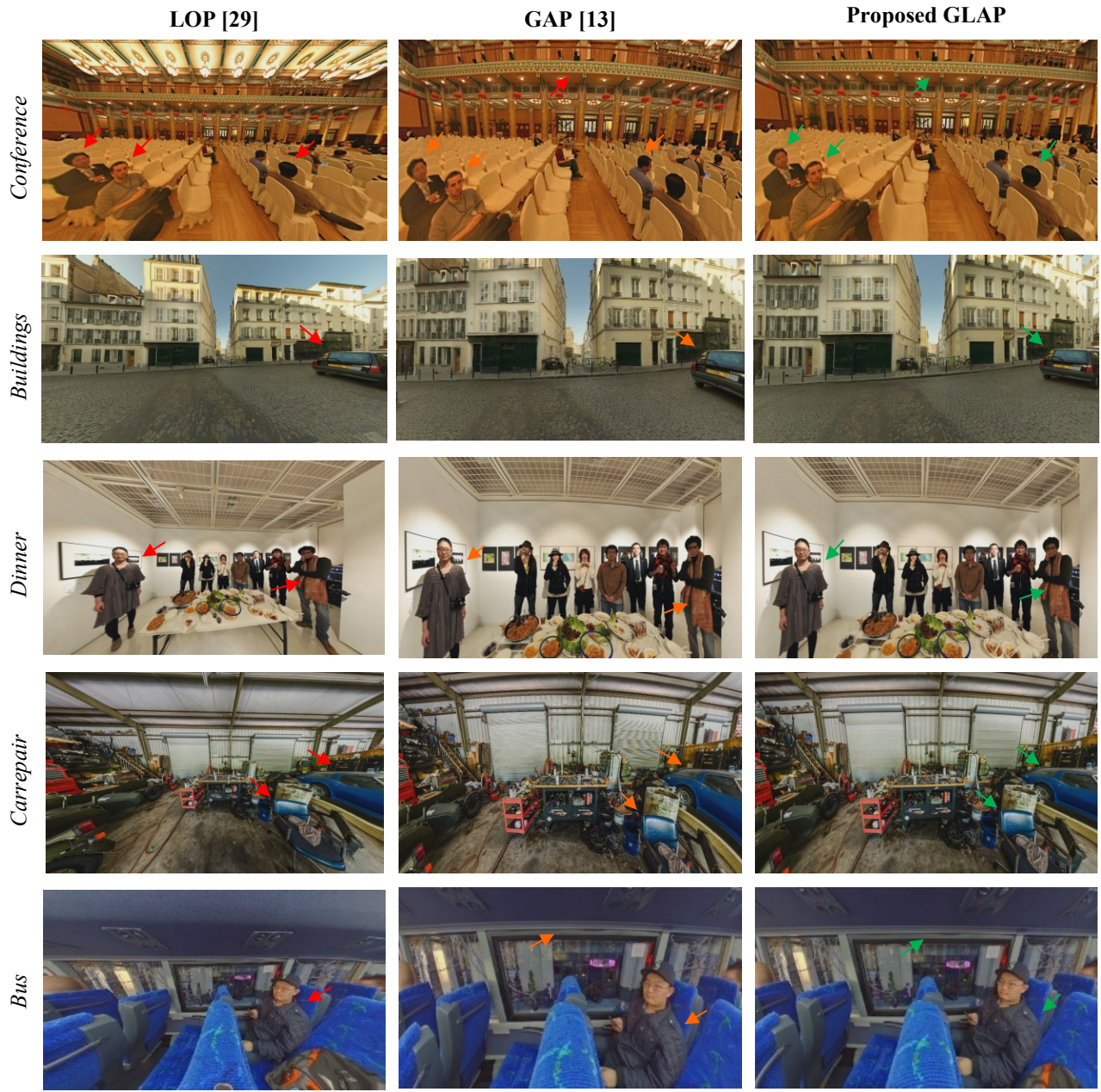

Figure 11: Viewport examples obtained with LOP from [29], GAP from [13], and proposed GLAP projection, using a HFoV of 150º.

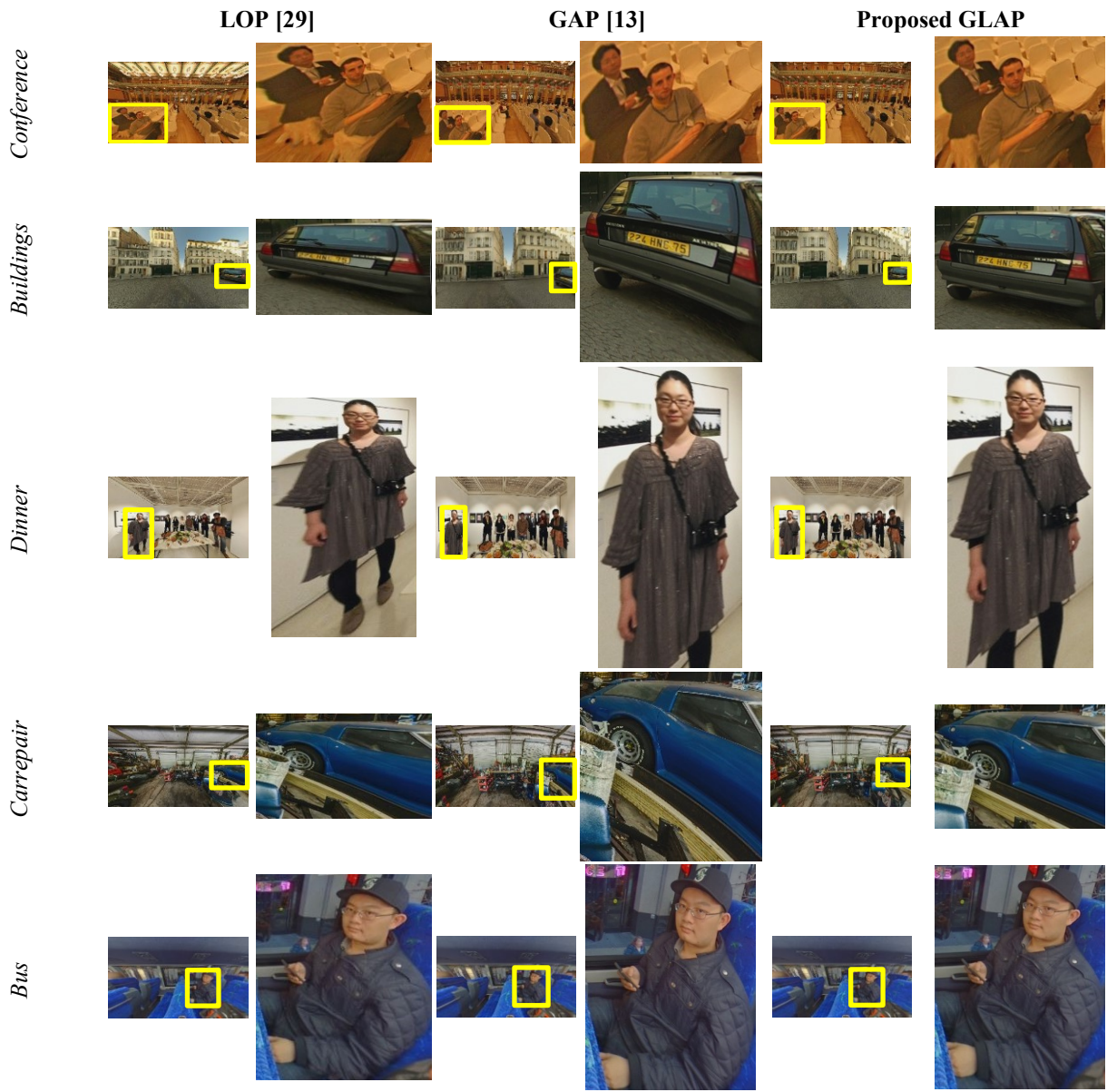

Figure 12: Comparing object conformality for viewport examples obtained with LOP from [29], GAP from [13], and proposed GLAP projection, using a HFoV of 150º.